\documentclass[doublecol]{epl2}
\usepackage{dcolumn}
\usepackage{bm}
\usepackage{latexsym}
  \usepackage{amsmath}
 \usepackage{amsthm}
 \usepackage{amsfonts}
 \usepackage{epsfig}
 \usepackage{graphics}
 \usepackage{dsfont}				
\usepackage{psfrag}
\usepackage{pstricks,pst-grad}

\title{Resonant tunneling-based spin ratchets}

\author{Matthias Scheid\thanks{\email{Matthias.Scheid@physik.uni-r.de}} \and Andreas Lassl \and Klaus Richter}
\shortauthor{M. Scheid \etal}
\institute{Institut f\"ur Theoretische Physik, Universit\"at Regensburg, 93040 Regensburg, Germany}

\abstract{
We outline a generic ratchet mechanism for creating directed spin-polarized currents 
in ac-driven double well or double dot structures by employing resonant spin
transfer through the system engineered by local external magnetic fields.
We show its applicability to semiconductor nanostructures 
by considering coherent transport through two coupled lateral quantum dots,
where the energy levels of the two dots exhibit opposite Zeeman spin splitting.
We perform numerical quantum mechanical calculations for the
$I$-$V$ characteristics of this system in the nonlinear regime, which requires
a self-consistent treatment of the charge redistribution due to the applied 
finite bias.
We show that this setting enables nonzero averaged net spin currents 
in the absence of net charge transport. 
}

\begin{document}

\maketitle

\section{Introduction}
The field of semiconductor spintronics has seen rapid progress lately, yet there
are still many obstacles on the way from fundamental research to operating
spin-based devices~\cite{Awschalom2007}. The creation of spin 
polarized currents is one basic requirement for the realization 
of semiconductor spintronics systems that share the prospect of being able to outperform
conventional electronics. Due to a better controlability and faster processing
times it is favorable to generate those currents by electrical means,
e.g. by the variation of (contact) voltages. 
Promising classes of devices include spin 
pumps~\cite{Mucciolo2002,Watson2003,Sharma2003,Governale2003}, 
spin rectification~\cite{Braunecker2007} and spin
ratchets~\cite{Scheid2006,Scheid2007,Scheid2007a,Smirnov2008,Smirnov2008a,Smirnov2009}. 
These proposals share the common idea to generate directed
spin currents, e.g.\ mediated by spin-orbit interaction, upon time variation
of external potentials. Here we focus on spin ratchets, a generalization of 
the particle quantum ratchet mechanism~\cite{Reimann1997,Linke1999,Haenggi2009}.
In such systems with broken spatial symmetry,
pure spin currents are generated by means of an ac-driving with no net average 
bias. This idea has been put forward for both, nonlinearly driven coherent conductors
\cite{Scheid2006,Scheid2007,Scheid2007a}, as well as conductors in the dissipative 
regime, where Brownian particle motion is converted into directed spin currents
\cite{Smirnov2008,Smirnov2008a,Smirnov2009}. While a net spin current could be shown to exist
for the different settings, its magnitude is difficult to predict and an optimization
towards larger spin currents is often not evident.  

Here we propose another, generic, spin ratchet mechanism that is based on coherent
{\em resonant} charge and spin transfer. It thereby leads to larger and controllable 
output and can be implemented in a variety of systems. 
Moreover, since the ratchet spin currents require operation under
nonequilibrium conditions and since the spin currents can usually be enhanced for strong
ac-bias, we employ a fully self-consistent treatment of the electrostatics for
our quantum transport calculations in the nonequilibrium regime. 
This involves a self-consistent determination of the voltage drop across the 
ratchet, which has been approximated so far only by simple heuristic models 
\cite{Scheid2006,Scheid2007,Scheid2007a}.

%
%
%
\begin{figure*}[t]
	\centering 
	\includegraphics[width=0.23\linewidth]{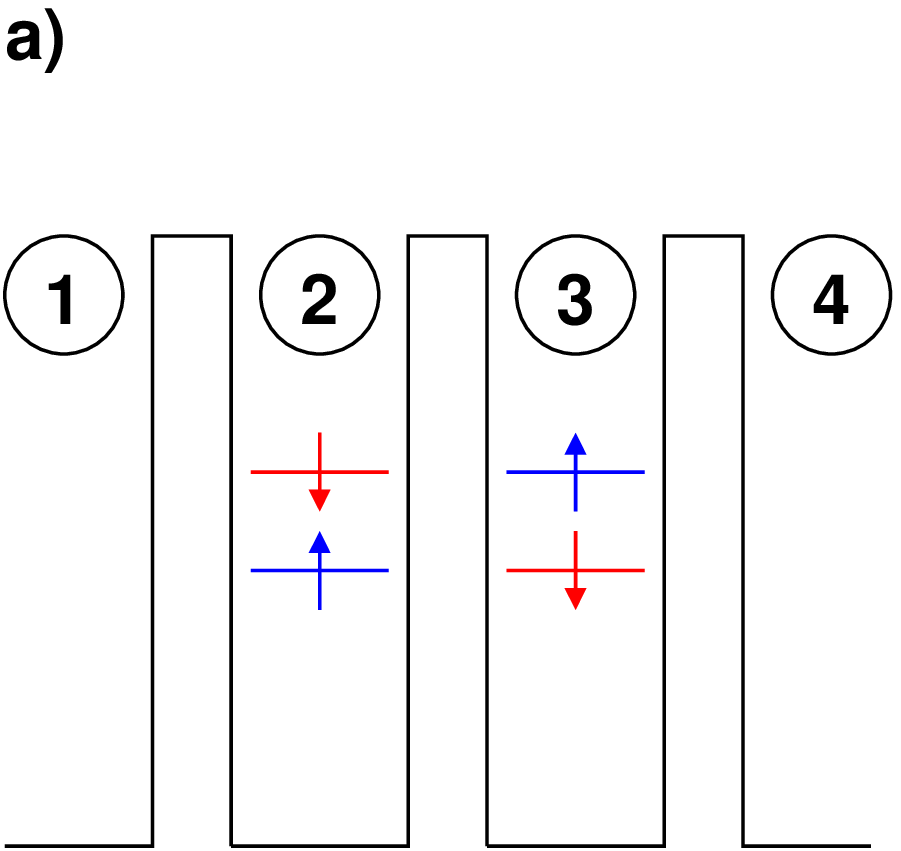}
	\hspace*{0.06\linewidth}
	\includegraphics[width=0.24\linewidth]{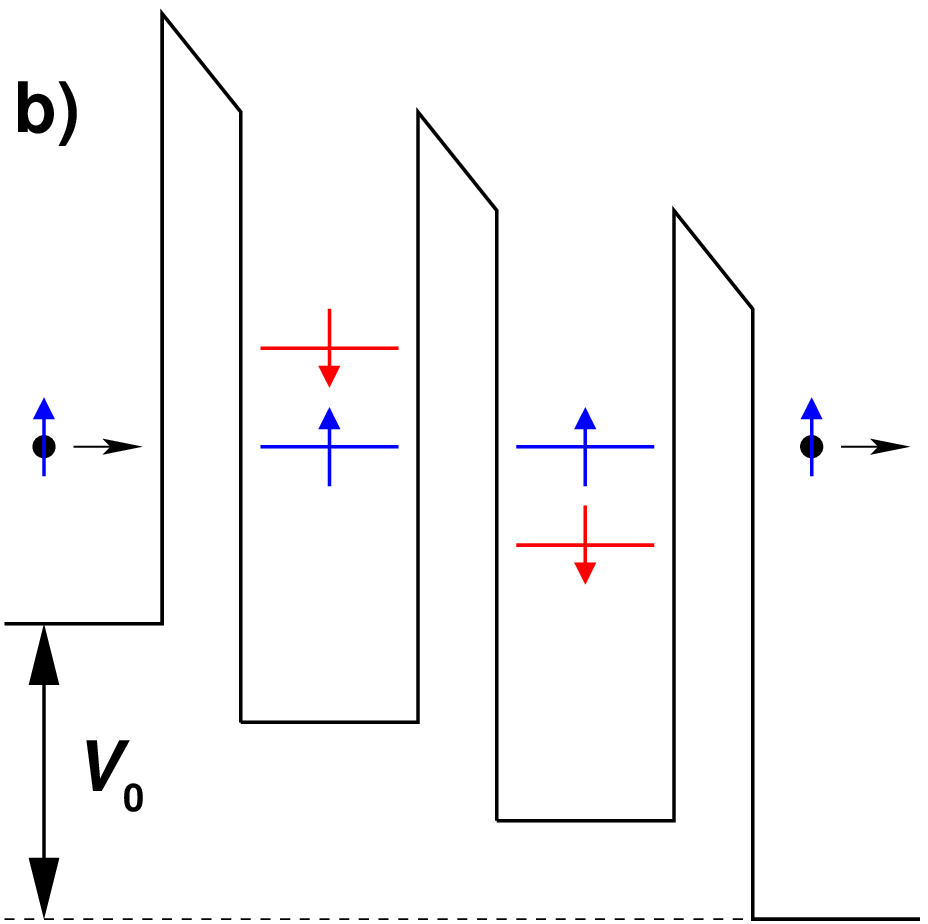}
	\hspace*{0.06\linewidth}
	\includegraphics[width=0.24\linewidth]{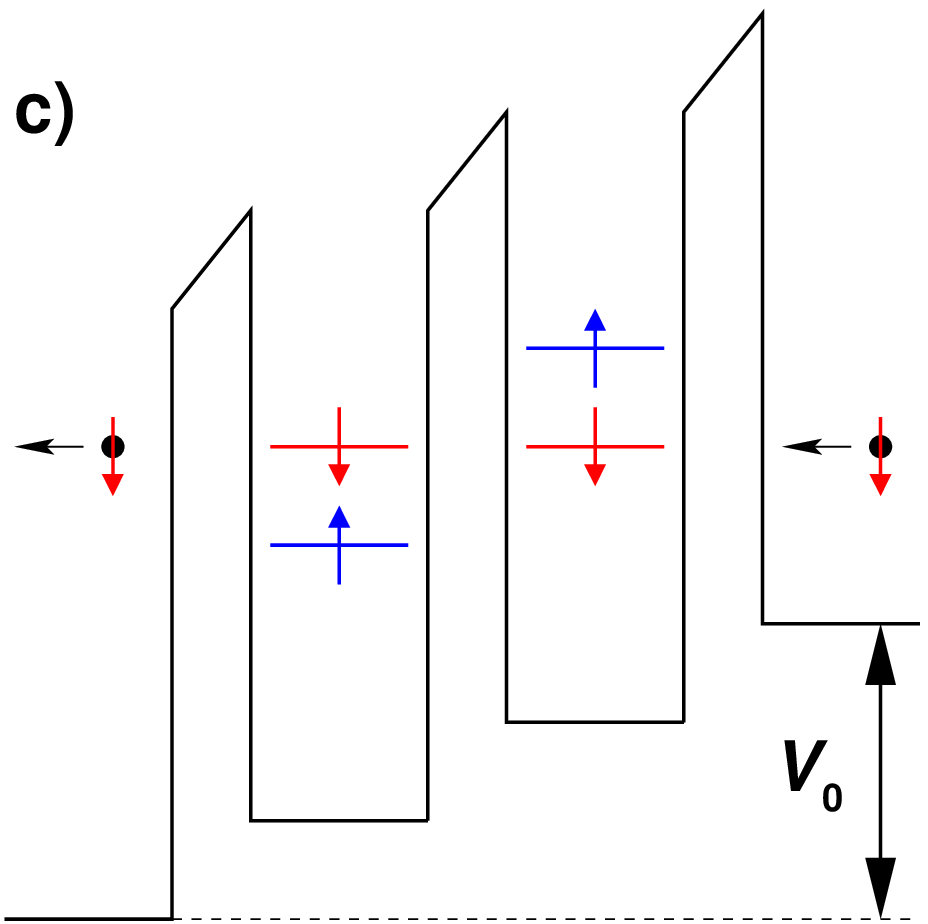}
	\caption{(\emph{Color online}) Illustration of the working principle of 
	a resonant spin ratchet. Panel a): system devided into four regions R1 to R4 
	through three barriers. A magnetic field, oriented in opposite directions in 
	regions R2 and R3, splits the levels of spin-up and spin-down electrons
	(one representative level shown). Panels b,c): Upon application of a 
	positive or negative bias voltage, $\pm V_0$, the transmission probability is
	resonantly enhanced for spin-up or spin-down electrons, respectively. 
	\label{Sketch}}
\end{figure*}
%
%
%

After outlining the general working principle we focus on a setup invoking
resonant tunneling through two quantum dots (QD) in a two-dimensional electron
gas (2DEG). In the literature double QD systems have already been proposed as 
spin filters~\cite{Cota2003,Cota2005} 
or sources for pure spin currents~\cite{Sun2003}. However, these proposals
are based on QDs in the Coulomb blockade regime, whereas the double QDs 
considered here are strong coupled to the leads, i.e.\ transport is fully coherent
and the conductance is larger. 

\section{Mechanism}
To illustrate the envisioned spin ratchet mechanism, let us first consider the
simplified one-dimensional potential model shown in Fig.~\ref{Sketch}a).  
Three electrostatic barriers divide the system into four regions (R1-R4). 
While the regions R1 and R4 support states with a continuous energy spectrum,
the regions R2 and R3, representing the double QD structure, accommodate
discrete resonant states due to the confinement imposed by the barriers. 
We further assume a magnetic field oriented in opposite directions in R2 and R3.
Thereby, the resonant energy levels in R2 and R3 are spin split due to the 
Zeeman coupling, however oppositely in both regions.
A finite bias voltage across the device shifts the energy levels in R2 with 
respect to those in R3. This enables one to bring energy levels of the same spin 
state in the two QDs into resonance at a specific bias voltage, see
Fig.~\ref{Sketch}b,c). 
For charge transport through the device, this results in an enhanced 
transmission of electrons of that specific spin state. Considering both forward 
and backward bias, it is obvious that the energy levels of different spin states 
can be brought into resonance for different signs of the bias voltage
(see Fig.~\ref{Sketch}b,c)).
Therefore, upon applying an ac-bias to the system, spin can be transported 
in the absence of a net charge current, as we will confirm below. 

\section{System and method}
\begin{figure}
	\centering
	\includegraphics[width=0.9\linewidth]{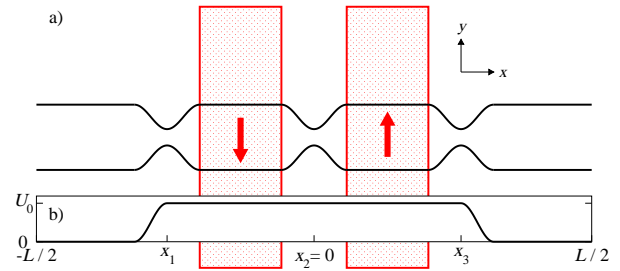}
	\caption{(\emph{Color online}) 
	a) Possible experimental realization of the principle mechanism for spin 
	current generation described in Fig.~\ref{Sketch}: 
	Two quantum dots are electrostatically defined (e.g.\ via side gates)
	within a quantum wire patterned on a 2DEG. The magnetic fringe fields of 
	two oppositely magnetized ferromagnetic stripes create the opposite Zeeman 
	splitting in the two quantum dots. b) Sketch of an additional electrostatic 
	potential, see text.
	\label{Setups}}
\end{figure}

In Fig.~\ref{Setups} we depict a possible experimental setup, which takes advantage of 
the principle just described and can be realized with present day material processing 
techniques. It is based on a quantum wire (QW) patterned on a 2DEG in the $(x,y)$-plane, 
see Fig.~\ref{Setups}a. 
Within this wire (in $x$-direction) which is connected to two non magnetic leads, 
two QDs possessing discrete energy levels are formed, e.g.\ via side gates. To 
realize the opposite Zeeman splitting inside these QDs, two ferromagnetic stripes (FMS)
with opposite in-plane magnetization $(\vec{M}=\pm M\hat{y})$ are patterned on top of 
the semiconductor heterostructure. The fringe fields of the FMS give rise to a 
non-uniform magnetic field $\vec{B}(x,y)$ in the plane of the 2DEG. We note that
the proposed setup is just one possible realization of the mechanism 
to generate pure spin currents outlined in Fig.~\ref{Sketch}. Alternatively, one could
for instance think of charge transport through resonant tunneling diodes, where the 
Zeeman splitting can be introduced, e.g., by layers of dilute magnetic 
semiconductors~\cite{Slobodskyy2003,Ertler2006}.
%
%

The quantum dynamics of electrons in the conductor is described by the 
single-particle Hamiltonian
%
%
\begin{equation}
\mathcal{H}=\frac{\Pi _x^2+\Pi _y^2}{2m^*}+ \frac{g^*\mu_\text{B}}{2} \vec{B}(x,y)\cdot\vec{\sigma }+V(x,y) \, ,
\end{equation}
%
where $m^*$ is the effective mass, $g^*$ the effective gyroscopic factor, 
$\mu_\text{B}$ the Bohr magneton, and 
$\vec{\sigma}$ is the vector of the Pauli spin operators.  
Orbital effects due to the magnetic field are accounted for by the vector
potential $\vec{A}(x,y)$, which enters the momenta $\Pi _i(x,y)=p_i-eA_i(x,y)$, 
while the Zeeman term $\frac{1}{2} g^*\mu_\text{B}\vec{B}(x,y)\cdot\vec{\sigma
}$ couples the spin degree of freedom to the external magnetic field
$\vec{B}(x,y)$ due to the FMS. For this setup $\vec{B}(x,y)$ can be evaluated 
-- along with $\vec{A}(x,y)$ -- using standard magnetostatics~\cite{Jackson1999}. 
The potential
$V(x,y)$ includes the lateral confining potential which forms the QW and QDs, 
a possible potential offset due to an additional gate voltage (see Fig.~\ref{Setups}b),
as well as the electrostatic potential due to an applied (driving) bias 
voltage $V_0$ between the left and right contact to which the QW is connected. The 
driving is assumed to be adiabatic.

We neglect inelastic processes and assume phase coherent electron transport. 
The bias $eV_0=\mu_\text{L}-\mu _\text{R}$ induces an electrical current, which 
we evaluate in the right lead. Specifically, the current of electrons with spin polarization 
$\sigma=\pm$ (with respect to a quantization axis in $y$ direction) 
can be written as 
\begin{equation}
\label{spin-current}
I_{\sigma}(V_0)  = \frac{e}{h}
\int_{E_\text{C}}^{\infty}  \text{d}E \, 
\Delta f(E;V_0) T_{\sigma}(E;V_0) \, .
\end{equation}
Here $E_\text{C}$ denotes the energy of the 
conduction band edge, and $\Delta f(E;V_0)= \left[ f(E,E_\text{F}+e\frac{V_0}{2}) -
f(E,E_\text{F}-e\frac{V_0}{2}) \right]$ is the difference between the 
Fermi functions in the two leads. In Eq.~(\ref{spin-current}), the quantum 
transmission probability for electrons with spin $\sigma$ is given by
\begin{equation}
T_\sigma (E;V_0)=\sum_n^{N_\text{L}}\sum_{n'}^{N_\text{R}} \sum_{\sigma '=\pm 1}
     \left| t_{n\sigma ,n'\sigma '}(E;V_0) \right| ^2,
\end{equation}
where $t_{n\sigma ,n'\sigma '}$ is the amplitude for transmission from the 
scattering state $(n'\sigma ')$ in the left lead into the scattering  state $(n\sigma )$ in the right 
lead, with the summations running over the $N_\text{L/R}$ open transversal channels
of the left/right lead. These amplitudes are evaluated by projecting 
the Green's function of the open system onto an appropriate set of asymptotic spinors 
defining incoming and outgoing channels. Making use of a real-space discretization 
of the Schr\"odinger equation~\cite{Ferry1997}, the calculation 
of the $\mathcal{S}$-matrix elements was made feasible by the implementation 
of a recursive algorithm for the calculation of 
Green's functions for spin-dependent transport~\cite{Lassl2007}.
\section{Self-consistent numerical procedure}
Since ratchet (spin) currents are expected for ac-driving 
with external bias voltages $V_0$ in the nonlinear regime, the  profile
of the electrostatic potential, dropping across the device, may 
play an important role for the working principle of the ratchet device,
and hence its treatment needs special care. In the following we outline our
approach to the non-equilibrium quantum transport problem including the
self-consistent determination of the electrostatic potential drop arising
from the charge rearrangement in the nonlinear bias regime. We consider
the classical electrostatic potential $V_{\rm es}(\vec{r})$ described by the 
Poisson equation, $\vec{\nabla}^2 V_{\rm es}(\vec{r}) =
e\rho(\vec{r})/(\varepsilon \varepsilon_0)$, with $\varepsilon =15.15$ for InAs. 
The charge density $\rho(\vec{r}) = -e[n(\vec{r})-n_d(\vec{r})]$ consists of
both the density distributions of the electrons, $n(\vec{r})$, and the donors,
$n_d(\vec{r})$. The latter is usually not known a priori, while the electron
density can be calculated from the lesser Green function 
$\mathcal{G}^<$, see e.g.~\cite{Ferry1997}:
\begin{equation}
  \label{eq:e_density}
  n(\vec{r}) = -\frac{\rm i}{2\pi} \int {\rm d}E\,
  \mathcal{G}^<(\vec{r},\vec{r};E) \, .
\end{equation}

In equilibrium the electrostatic potential is typically included in the
effective confinement potential $V = V_{\rm conf} + V_{\rm es}^0$ which we 
modelled by a hard-wall potential~\cite{Laux1988}.   Therefore, 
the equilibrium electrostatic potential, governed by 
$\vec{\nabla}^2 V_{\rm es}^0(\vec{r}) = 
-e^2\big[n_0(\vec{r})-n_d(\vec{r})\big] / (\varepsilon \varepsilon_0)$,
with $n_0(\vec{r})$ the electron density for zero-bias, need not be considered 
explicitly.
Thus, in the nonequilibrium situation with a finite source-drain bias, only the
change of the electrostatic potential $\delta V_{\rm es} = V_{\rm es}-V^0_{\rm
es}$ due to the charge rearrangement $\delta n = n-n_0$ of the electrons 
has to be included, similar as described in Ref.~\cite{Xue2002}.
Hence the Poisson equation
\begin{equation}
  \label{eq:poisson}
  \vec{\nabla}^2 \delta V_{\rm es}(\vec{r}) = 
  -\frac{e^2}{\varepsilon \varepsilon_0} \Big[n(\vec{r}) - n_0(\vec{r})\Big] .
\end{equation}
has to be solved.

To account for the influence of the leads it is convenient to write the
electrostatic potential as a sum, $\delta V_{\rm es} = V_{\rm bias} + V_{\rm
el}$ \cite{Xue2002}. Here, $V_{\rm bias}$ is the potential induced by the contacts 
ignoring the charges in the device. 
It is the solution of the Laplace equation $\vec{\nabla}^2 V_{\rm bias}(\vec{r})
= 0$ with boundary conditions $V_{\rm bias}(x=\pm L/2) = \pm eV_0/2$, where $L$ is
the distance between the contacts. 
For three-dimensional contacts, $V_{\rm bias}$ has the shape of a linear ramp.
The contribution of the charges inside the system is described by $V_{\rm el}$,
which is the solution of Eq.~(\ref{eq:poisson}) with the boundary conditions of 
$V_{\rm el}(x=\pm L/2) = 0$ at the interfaces of the contacts. 

If the electron density in the leads is substantially higher than in the device,
the potential drop is screened in the leads leading to a flat electrostatic
potential close to the contacts. 
Then, $n(\vec{r}\,') \approx n_0(\vec{r}\,')$ far from the device, and we can 
compute the electrostatic potential as 
\begin{equation}
  \label{eq:V_el}
  V_{\rm el}(\vec{r}) = \frac{e^2}{4\pi\varepsilon \varepsilon_0} \int {\rm d}^2
  r' \frac{n(\vec{r}\,') - n_0(\vec{r}\,')}{|\vec{r}-\vec{r}\,'|},
\end{equation}
which is the solution of the Poisson equation for a vanishing potential at
infinity. 

For the self-consistent solution of the transport problem 
we compute from Eq.~(\ref{eq:e_density})the
electron density $n(\vec{r})$ for a given electrostatic potential and obtain
an improved profile for the electrostatic potential from $n(\vec{r})$.
It is known, however, that the straightforward iteration between
Eqs.~(\ref{eq:e_density}) and (\ref{eq:V_el}) typically does not converge
\cite{Trellakis1997}. 
Instead we adapted the Newton-Raphson method introduced in
Refs.~\cite{Trellakis1997, Lake1997} in order to evaluate the electrostatic potential
by means of Eq.~(\ref{eq:V_el}), which significantly improves the convergence behavior
of the self-consistent computation scheme.
After calculating the self-consistent electrostatic potential for each value of
the source-drain voltage we are able to determine the transport properties of
the considered ratchet device using Eq.~(\ref{spin-current}).

\section{Numerical results}
%
%
\begin{figure}[t]
	\centering
	\includegraphics[width=0.8\linewidth]{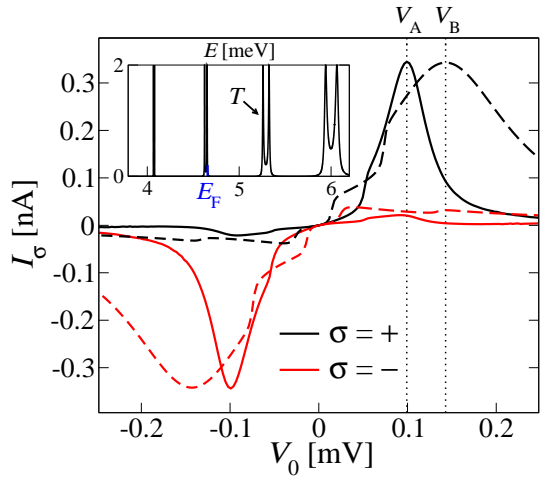}
  \caption{(\emph{Color online}) Spin resolved currents $I_\sigma$ for 
  spin-up ($\sigma=+$, black curve) and spin-down ($\sigma=-$, red curve) 
  as a function of applied bias voltage $V_0$ for an InAs-based double dot 
  device (Fig.~\ref{Setups}) at Fermi energy $E_\text{F}=4.66$ meV.
  The solid (dashed) lines denote the calculation
  for the self-consistent (model) electrostatic potential drop. 
  Inset: Transmission $T(E,V_0=0)$ through the same system
   at zero magnetic field in linear response.}
  \label{fig:spin_current}
\end{figure}
%
%
%
In the inset of Fig.~\ref{fig:spin_current} we show the transmission
$T(E;0)$ in the linear-response regime, $V_0 \rightarrow 0$, 
at zero magnetic field for the device depicted in Fig.~\ref{Setups}. The
system's geometry parameters (specified in detail below) and energies are chosen such
that the leads carry 3 to 4 transversal modes, 
while the point contacts only allow for (resonant) tunneling. 
Thus, we find several sharp transmission peaks corresponding to
the discrete resonant energy levels of the two QDs. The peaks appear as
doublets due to the inter-dot tunnel splitting.\\
Upon applying a voltage we expect an asymmetric \mbox{$I$-$V$} 
characteristic for the spin resolved currents for Fermi 
energies close to those resonant energy levels. 
To conduct explicit calculations for the spin current of the device 
shown in Fig.~\ref{Setups} we have to fix several parameters. Taking InAs as 
the material where the 2DEG is built from, we have $m^*=0.024 m_0$ and $g^*=15$. 
The width of the QW is chosen to be $W=200\mathrm{nm}$, while at the point contacts, 
which are separated by $450\mathrm{nm}$, the width narrows down to $70\mathrm{nm}$. The two 
FMS of identical size have dimensions $x_0=250\mathrm{nm}$, $y_0=800\mathrm{nm}$, 
$z_0=250\mathrm{nm}$ and are centered on top of the two point contacts at a 
distance $50\mathrm{nm}$ above the 2DEG. For the magnetization of the 
stripes we chose $\mu _0M=2.5\mathrm{T}$, a value well in 
reach using e.g. Dysprosium~\cite{Uzur2004}.
To allow for a high electron density in the leads in order to screen the
electrostatic potential drop and to achieve a flat electrostatic potential 
in R1 and R4, we include an additional confinement
potential $U_0 = 2.92 \rm meV$ as shown in Fig.~\ref{Setups}b). 

%
\begin{figure}
\begin{center}
	\includegraphics[width=0.8\linewidth]{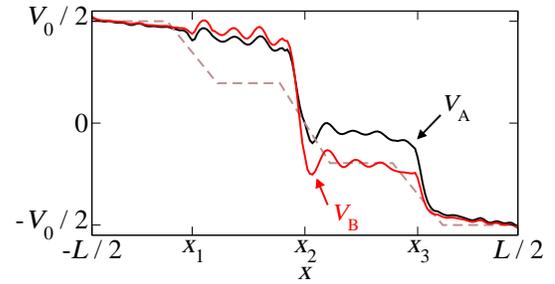}
  \caption{(\emph{Color online}) Full lines: Self-consistent electrostatic potential 
  drop $V_{\rm es}(x)$ for the two quantum dots bounded by points contacts at
  $x=x_1,x_2,x_3$ and averaged over the transverse direction. It is shown for 
  bias voltages $V_\text{A/B}$ as indicated in Fig.~\ref{fig:spin_current}. 
  The dashed line 
  corresponds to the heuristic voltage drop model (underlying the dashed spin
  current curves in Fig.~\ref{fig:spin_current}).}
  \label{fig:V_es}
\end{center}
\end{figure}
%

The spin-resolved $I$-$V$ characteristics which are obtained with the described procedure for
the Fermi energy at the second resonant energy level, $E_\text{F}=4.66$ meV
(see inset of Fig.~\ref{fig:spin_current}), are shown in Fig.~\ref{fig:spin_current} 
as solid lines (black and red line for spin-up and down, respectively). There we find
that the device indeed acts as a spin selective current rectifier with dominant $(>90\%)$
spin-up polarization for positive and spin-down polarization for negative bias voltages. 
This implies that when applying an adiabatically varying unbiased ac-voltage 
to the contacts the system only transports spins, yet no net charge, since the total
charge currents average to zero. The pronounced maximum (minimum) at $\pm V_\text{A}$
reflects level alignment, i.e. the resonance condition sketched in Fig.~\ref{Sketch}.
For comparison the dashed lines in Fig.~\ref{fig:spin_current} represent the 
corresponding $I_\sigma$-$V$ curves arising from the aforementioned heuristic 
voltage drop model. It assumes a local voltage drop ($V_0 /3$) at each of the 
three point contacts, i.e.\ the regions of maximum resistance. 
We find the same qualitative overall behavior for the spin currents, 
while the maximum of the resonant 
tunneling current is reached at a higher value, $V_0=V_\text{B}$ 
with $eV_\text{B}$ being approximately three times the Zeeman splitting 
of the energy levels of the QD, as expected for this particular model. 
We performed further calculations (not presented here) which show that the 
degree of spin polarization, i.e. the efficiency of the device, increases with 
the Zeeman splitting in regions R2 and R3, due to the stronger separation of 
the spin $- \sigma$ energy levels at resonance for spin $\sigma$. 
Accordingly, the optimal working condition, i.e. levels for one spin species at 
resonance, is then reached at higher voltages.

Since the amount of voltage that drops at the central point contact 
determines the alignment of the energy levels of the two QDs, in Fig.~\ref{fig:V_es} 
we study the profile of the self-consistent electrostatic potential, $V_{\rm es}(x)$,
along the transport direction for applied bias $V_0=V_\text{A}$ (full black line)
and $V_\text{B}$ (full red line) with  $V_\text{A/B}$ marked in
Fig.~\ref{fig:spin_current}. We see that due to the point contacts a step structure 
emerges, with its exact form depending on $V_0$ and $E_\text{F}$.
Compared to the heuristic voltage drop model (dashed line), we find a stronger
voltage drop at the central point contact. This explains that in 
Fig.~\ref{fig:spin_current} the current maximum is reached at lower bias voltages
and the currents are significantly smaller at high bias for the self-consistent 
current calculation.  Still, in view of Figs.~\ref{fig:spin_current} 
and~\ref{fig:V_es}, we can conclude that the heuristic voltage drop model constitutes 
a fair approximation to $V_{\rm es}$ for the system considered.

\section{Summary}
In conclusion we have presented a generic mechanism to produce pure spin currents in nanostructures by 
applying an ac electrical bias. 
It is based on resonant transfer of spin-up and -down electrons in opposite directions and gives rise 
to pure ratchet type spin currents upon driving. Our results are based on self-consistent Keldysh Greens 
function transport in order to adequately describe the nonequilibrium condictions under which the 
system works.

\acknowledgements
We acknowledge useful conversations with C.~Ertler, M.~Wimmer and D.~Bercioux. The work was funded by the
{\em Deutsche Forschungsgemeinschaft} within SFB 689. MS acknowledges additional funding from
{\em Studienstiftung des Deutschen Volkes}.
%
%
%
%
%
%

\end{document}